\documentclass{article}
\usepackage[fontsize=9.4pt]{fontsize}
\usepackage{spconf,amsmath,graphicx,amsfonts,amssymb,booktabs,siunitx}

\title{Sinusoidal Frequency Estimation by Gradient Descent}
\name{Ben Hayes\thanks{Ben Hayes is
supported by UK Research and Innovation [grant number
EP/S022694/1]. This research utilised Queen Mary's Apocrita HPC facility, supported by QMUL Research-IT. \url{http://doi.org/10.5281/zenodo.438045}}, Charalampos Saitis, Gy{\"o}rgy Fazekas}
\address{Centre for Digital Music, Queen Mary University of London}
\begin{document}
%\ninept
%
% \vspace{-12pt}
\maketitle
\begin{abstract}
Sinusoidal parameter estimation is a fundamental task in applications from spectral analysis to time-series forecasting.
Estimating the sinusoidal frequency parameter by gradient descent is, however, often impossible as the error function is non-convex and densely populated with local minima.
The growing family of differentiable signal processing methods has therefore been unable to tune the frequency of oscillatory components, preventing their use in a broad range of applications.
This work presents a technique for joint sinusoidal frequency and amplitude estimation using the Wirtinger derivatives of a complex exponential surrogate and any first order gradient-based optimizer, enabling end-to-end training of neural network controllers for unconstrained sinusoidal models.
% This work presents a technique for sinusoidal frequency estimation by gradient descent under specific conditions, and proposes a direction for future work towards a general solution.
\end{abstract}
% EMD, meta-loss, cumsum gradient, IIR surrogate
%
%
% \vspace{-1mm}
\section{Introduction}
\label{sec:intro}

% \section{Background}
% \label{sec:related}

Estimating sinusoidal parameters from a signal is a crucial step in numerous signal processing algorithms, and a wealth of techniques have been proposed in both the single and multiple sinusoid formulations. Most seek the maximum likelihood (ML) estimate of sinusoidal model parameters in the presence of white Gaussian noise, the statistical properties of which are well established \cite{kay_fundamentals_1993}.

Estimators of sinusoidal frequency must circumvent the non-linearity of the model and non-convexity of the corresponding objective as a function of the frequency parameter.
The most common approach is thus to apply a multi-stage algorithm in which an initial frequency estimate is obtained through search heuristics \cite{stoica_maximum_1989, rife_single_1974, abatzoglou_fast_1985}, spectral peak interpolation \cite{rife_multiple_1976}, discrete-time Fourier transform (DTFT) decorrelation \cite{abeysekera_least-squares_2018}, or other procedures \cite{parthasarathy_maximum-likelihood_1985, tseng_low-complexity_2022, vishnu_improved_2020}, and then refined using an optimization method.
Alternate approaches include iteratively updating a model-based relaxation of the problem \cite{bresler_exact_1986, tseng_low-complexity_2022},  linearizing the problem using delay operators \cite{gromov_first-order_2017, vediakova_finite_2021}, or defining a surrogate model where an equivalence can be drawn between solutions \cite{kumaresan_algorithm_1986}.

Such methods achieve accurate estimates but are unsuitable for use in the context of end-to-end models fit by gradient descent, where integrating derivative-free operations or complex heuristics is challenging and often unstable.
In particular, the recent proliferation of models applying \textit{differentiable digital signal processing} (DDSP) \cite{engel_ddsp_2020} --
% Such a solution to the frequency estimation problem would greatly benefit the \textit{differentiable digital signal processing} (DDSP) \cite{engel_ddsp_2020} 
a family of techniques which allow neural networks to directly control digital signal processors -- highlights the need for a method for sinusoidal frequency estimation by gradient descent.

Applications of DDSP have included providing high level controls for harmonic-plus-noise synthesizers \cite{engel_ddsp_2020}, controlling digital synthesis methods with neural networks \cite{hayes_neural_2021, caspe_ddx7_2022}, modelling \cite{nercessian_lightweight_2021} and controlling \cite{steinmetz_style_2022} audio effects and direct filter design \cite{colonel_direct_2022}. 
Yet, despite success at these complex tasks, DDSP-based models have so far been unable to predict sinusoidal frequency parameters. Aspects of the problem have been acknowledged in the literature. Turian \& Henry \cite{turian_im_2020} showed that frequency domain distances lack a stable and informative frequency gradient, whilst Engel et al. \cite{engel_self-supervised_2020} used a parameter regression pretraining scheme to circumvent issues with local minima when optimizing sinusoidal frequencies. Caspe et al. \cite{caspe_ddx7_2022} similarly note that gradient descent fails to tune the modulation frequencies of a differentiable FM synthesizer due to ripple in the error function.

% Here, we argue the frequency estimation problem can be traced to three root causes: (i) local minima caused by spectral leakage, (ii) a lack of informative frequency gradient, (iii) interference and instability caused by aggregating gradients across time.
% We argue that real sinusoidal gradients are unsuitable for gradient descent, and instead 
In this work, we propose a simple surrogate to the sinusoidal oscillator with gradients that allow first-order gradient based optimization.
With this approach, we take a first step towards end-to-end learning of neural network controllers for a broader family of differentiable audio synthesizers and signal processors.
% providing estimates of sinusoidal parameters in both single and multiple sinusoid settings.

%, hayes_neural_2021, colonel_direct_2022, caspe_ddx7_2022}

% DDSP
% sinusoidal parameter estimation
% - mention older literature ? julius smith spectral analysis, etc. maybe reference sinusoidal least squares and requirement for a good initial estimate?
% - oscillator tuning (mention turian, inv-ddsp, ddx7, bechtle)
% oscillator tuning problem formally?

\vspace{-1mm}
\section{Sinusoidal Frequency Estimation}
\label{sec:sinusoidal}
\begin{figure*}
    \centering
    \includegraphics[width=\textwidth]{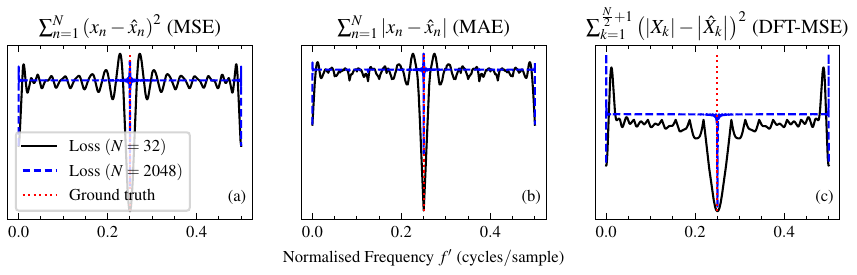}
    \caption{The (a) mean squared error, (b) mean absolute error, and (c) DFT modulus mean squared error loss computed between two sinusoidal signals ($N=32$ and $N=2048$) as a function of the predicted frequency. Spectral leakage results in local minima across the full parameter range at lower $N$, while higher $N$ reveals the lack of informative frequency gradient, preventing effective first order optimization.}
    \label{fig:mse_leakage}
    \vspace{-4mm}
\end{figure*}

We are concerned with modelling the class of discrete-time signals that can be expressed as:

\begin{equation}
\label{eqn:model}
    x_n = v_n + \sum_{k\in K} \alpha_k \cos \left(\omega_k n + \phi_k\right),
\end{equation}

\noindent where $v_n \sim \mathcal{N}\left(0, \sigma^2\right)$, and $\alpha_k, \omega_k, \phi_k$ are the amplitude, frequency, and phase parameters, respectively, of unordered sinusoidal components with index set $K\subseteq \mathbb{N}$.
% \footnote{No ordering of components should be inferred from the indexing -- in fact, the model is permutation invariant in the index set $K$}. %For convenience, parameters are collectively referred to as $\theta$.
Following the standard ML derivations, finding estimates $\hat{\alpha}_k$, $\hat{\omega}_k$, $\hat{\phi}_k$ is equivalent to minimizing the mean squared error of the model.
In many applications of machine learning to audio, we are concerned with other formulations of the error.
These can be accounted for by expressing the likelihood in terms of other signal representations, such as the discrete Fourier transform (DFT).
% Mean absolute error can be derived by instead assuming a Laplace distribution for $v_n$.
% 
% :
% 
% \begin{equation}
%     \mathcal{L}\left(\hat{\theta}; \mathbf{x} \right)
%     \propto \exp\left[-\frac{\sigma^2}{2}\left(\mathbf{x} - \hat{\mathbf{x}}\right)^T \left(\mathbf{x} - \hat{\mathbf{x}}\right)\right] 
% \end{equation}
% 
% where $\hat{\mathbf{x}}$ is obtained by applying Eqn. \ref{eqn:model} to parameter estimates $\hat{\theta}$. This is equivalent to minimising the mean squared error:
% \begin{equation}
% \label{eqn:objective}
%     \theta^\ast = \arg\min_{\hat{\theta}} \sum_{n=1}^{N} \left(x_n - \hat{x}_n\right)^2
% \end{equation}
% where $N$ is the length of the signals under analysis.

It is well established that when $\omega_k$ and $\phi_k$ are known, this problem is linear in $\alpha_k$ \cite{smith_spectral_2011} -- a property which, for example, allows DDSP models to directly predict harmonic amplitudes \cite{engel_ddsp_2020}. When $\phi$ is unknown, an optimal estimate can be found by evaluating the DTFT at the known frequencies $\omega_k$. In the case where frequency is unknown, however, the optimization problem is more challenging. Substituting Eqn. \ref{eqn:model} into the mean squared error, for example, it is clear that the minimization objective is highly non-convex (as illustrated in Fig. \ref{fig:mse_leakage}), consisting of a sum of second order intermodulation products.

% \begin{align}
% \begin{split}
% \label{eqn:objective}
%     \mathcal{L_\text{MSE}} =
%     \frac{1}{N}\sum_{n=1}^N \Biggl[
%     v_n &+ \sum_{k\in K} \alpha_k \cos\left(\omega_k n + \phi_k\right)\\
%     &- \hat{\alpha}_k \cos\left(\hat{\omega}_k n + \hat{\phi}_k \right) \Biggr]^2
% \end{split}
% \end{align}
% \noindent where $N$ is the length of the signal.
As only the global minimum\footnote{Strictly, there exist $K!$ such minima as all permutations of sinusoidal components are equivalent under summation. However, we concern ourselves here solely with arriving at \textit{any} of them, leaving study of the symmetries of the loss surface to future work.} represents a viable model fit, such a loss surface is problematic for gradient based optimizers: unless parameter estimates are initialized in the main basin of the function, optimization will converge on an incorrect solution.
Further, as Turian \& Henry \cite{turian_im_2020} note, the gradient of audio loss functions is generally uninformative with respect to frequency.
This is intuitive, considering the orthogonality of sinusoids on an infinite time horizon.
As $N\to\infty$, the ripple reduces in magnitude until, in the limit, the loss is zero when $\hat{\omega}=\omega$ and constant for all other $\hat{\omega}$.
The effect of increasing $N$ is illustrated in blue in Fig. \ref{fig:mse_leakage}.

\vspace{-1mm}
\subsection{Complex exponential surrogate}

\begin{figure}[t]
    \centering
    \includegraphics[width=\columnwidth]{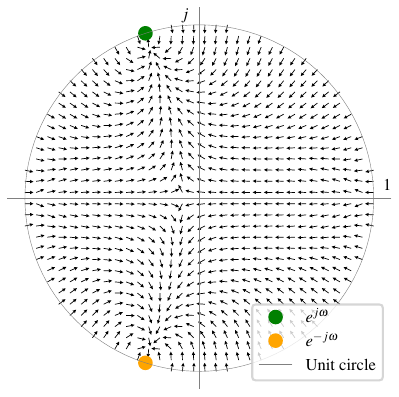}
    \caption{Directions of negative Wirtinger conjugate derivatives $\frac{\partial \mathcal{L}_\text{MSE}}{\partial \bar{z}}$ displayed on the complex plane, where $\mathcal{L}_\text{MSE}$ is the mean-squared error taken with a single sinusoid of target frequency $\omega$. Arrows point in the direction of steepest descent. Our approach uses the Wirtinger differential operator with a complex exponential as a surrogate for a sinusoidal oscillator to allow frequency estimation in a differentiable signal processing framework.}
    \label{fig:complex_field}
\end{figure}

Our proposed technique circumvents these issues by defining a surrogate for a differentiable sinusoidal model.
The surrogate produces an exponentially decaying sinusoid as the real part of an exponentiated complex number:

\begin{align}
    \mathfrak{s}_n(z_k) \triangleq
    \mathfrak{Re}\left(z_k^{n}\right)
    &= \left| z_k \right|^n \cos n \angle z_k \\\nonumber
    % \mathfrak{Im}\left(z^n\right)
    % &= \left| z \right|^n \sin n \angle z \nonumber
\end{align}

\noindent where $Z = \left\{z_k \in \mathbb{C} \mid k \in K \right\}$ is a set of specific surrogate parameters with index set $K$.
As the surrogate maps $\mathfrak{s}_n\colon \mathbb{C} \to \mathbb{R}$ it does not have a complex derivative.
However, its partial derivatives can be computed using Wirtinger's calculus.
% which, amongst other properties, allows the gradient of a real valued function of a complex variable to be evaluated in the complex plane.
A detailed explanation of these operators is beyond the scope of this paper, but we refer the interested reader to the work of Kreuz-Delgado \cite{kreutz-delgado_complex_2009} for an introduction.
For present purposes, the conjugate Wirtinger derivative of the surrogate is:

% we can apply the Wirtinger differential operators to obtain partial derivatives in the complex plane:
% Updates to $z$ are then given by:
% % If $z = x + jy$, we have:

% \begin{equation}
%     z_{t+1} = z_t - \eta\frac{\partial \mathcal{L}}{\partial \mathfrak{Im}\left(z_t^n\right)}\frac{\partial \mathfrak{Im}\left(z_t^n\right)}{\partial \bar{z}_t},
% \end{equation}

% \noindent where the conjugate Wirtinger derivative is given by: 

\begin{align}\label{eqn:wirt_grad}
\begin{split}
% \frac{\partial }{\partial z}\mathfrak{s}_n(z)
% &= \frac{1}{2}\left(\frac{\partial}{\partial x}-j\frac{\partial}{\partial y}\right)\mathfrak{s}_n(z)
% = \frac{n}{2}z^{n-1}\\
\frac{\partial }{\partial \bar{z}}\mathfrak{s}_n(z) 
&= \frac{1}{2}\left(\frac{\partial}{\partial x}+j\frac{\partial}{\partial y}\right)\mathfrak{s}_n(z)
= \frac{n}{2}\bar{z}^{n-1}\\
% \\
% &= \frac{n}{2}\left(z^{n-1}+\bar{z}^{n+1}\right)
% \frac{\partial }{\partial z}\mathfrak{Im}\left(z^n\right) &=
% \frac{1}{2}\left(\frac{\partial}{\partial x} - j\frac{\partial}{\partial y}\right)\mathfrak{Im}\left(z^n\right) \\
% &= \frac{n}{2}\left(z^{n-1}+\bar{z}^{n+1}\right)
\end{split}
\end{align}

\noindent for $z=x+jy$.
Where $\mathcal{L}$ is the loss between a signal produced by $\mathfrak{s}$ and a target, $-\frac{\partial \mathcal{L}}{\partial \bar{z}}$ is then the direction of steepest descent \cite{kreutz-delgado_complex_2009}.
% Full derivations of these gradients can be found in the online supplementary material\footnote{Add supplementary material URL}.
% When summed across the length of an analysis window, the negative conjugate Wirtinger derivatives of the loss function with respect to the surrogate parameter $z$ lead to a global minimum.
Unlike the frequency parameter of a real sinusoid, the surrogate parameter allows both the frequency and amplitude decay of the signal to be varied.
As illustrated in Fig. \ref{fig:complex_field}, an optimizer can thus move the parameter inside the unit circle, creating an exponential amplitude decay, before moving it back out at the correct angle from the real line.

The frequency estimate represented by the minimum is given by the complex argument of the surrogate parameter $\omega^\ast_k = \angle z_k^\ast$, where $Z^\ast = \left\{z_k^\ast \in \mathbb{C} \mid k \in K \right\}$ minimizes the loss:

\begin{equation}\label{eqn:freq_estimator}
    % Z^\ast = \arg\min_{Z} \sum_{n=1}^N \left(x_n - \sum_{k\in K} \mathfrak{s}_n\left(z_k\right)\right)^2
    Z^\ast = \arg\min_{Z} \sum_{n=1}^N \left(x_n - \sum_{k\in K} \mathfrak{s}_n\left(z_k\right)\right)^2
\end{equation}

\noindent It follows that when target amplitudes $\alpha_k=1$, this gives $|z^\ast_k|=1$ and $\angle z^\ast_k=\omega_k$, which is equivalent to the ML estimate for frequency.
In other cases, a linear amplitude parameter $\hat{\alpha}_k$ can be introduced to the surrogate, i.e. $\hat{\alpha}_k \mathfrak{s}_n\left(z_k\right)$.
In this way, $\omega_k^\ast$ can give the ML estimate when $\alpha_k \neq 1$ and $\alpha_k$ is known, by simply setting $\hat{\alpha_k}=\alpha_k$.
In the case where target amplitudes are unknown, $\hat{\alpha}_k$ can be learned jointly with surrogate parameters.

% and $\alpha_k \neq 1$, however, the estimator in Eqn \ref{eqn:freq_estimator} is biased.

% The Jacobian of $\mathbf{z}$ is given by :

% \begin{equation}
%     \frac{\partial \mathcal{L}}{\partial z_i} =
%     - \sum_{n=1}^N
%     n
%     \left(
%     z_i^{n-1}+z_i^{-n-1}
%     \right)
%     \left(
%     x_n
%     -
%     \sum_{k=1}^K
%     \mathfrak{Im}\left(z_k^n\right)
%     \right)
% \end{equation}

% \begin{equation}
%     \mathfrak{Im}\left(z^n\right) =
%     \frac{1}{2}
%     \left(
%     \left(x+jy\right)^n
%     +\left(x-jy\right)^n
%     \right)
% \end{equation}

% \begin{equation}
%     \mathcal{L} =
%     \sum_{n=1}^N \Biggl(
%     v_n + \alpha \sin\left(\omega n\right)
%     - \mathfrak{Im}\left(z^n\right) \Biggr)^2
% \end{equation}

% \subsubsection{Frequency estimation}

\vspace{-1mm}
\subsubsection{Amplitude estimation}

Even when an amplitude coefficient is jointly learned, the complex exponential surrogate and the standard sinusoid may differ in the evolution of their amplitudes across time.
% The sinusoidal model we aim to fit (Eqn \ref{eqn:model}) specifies a single steady amplitude for each component, whereas the surrogate's amplitude may decay exponentially over time to zero if $|z_k|<1$ or explode to infinity if $|z_k|>1$.
% The modulus of a surrogate estimate $|z^\ast_k|$ (or $\hat{\alpha}_k|z^\ast_k|$ where an amplitude factor is learned) is thus usually a poor estimator of the target amplitude.
However, assuming $z^\ast_k$ minimizes the least squares objective, we can recover an amplitude estimate by solving the opposite least squares problem.
% % In the single sinusoid case, we can derive from the objective function a transformation of $|z|$ that recovers the target amplitude. To do so, we assume that $\angle z = \omega$ and find the minimum of the error with respect to the amplitude.
% % For time-domain mean-squared error, this gives:
% % \begin{equation}
% %     \alpha^\ast_\text{MSE} = \frac{\sum_{n=1}^N n|z|^{2n-1}\sin^2 \omega n}{\sum_{n=1}^N n|z|^{n-1}\sin^2 \omega n}
% % \end{equation}
% % In the single sinusoid case, if $|z|$ minimises the objective function, this provides an exact estimate of $\alpha$.
Specifically, to recover amplitudes from a surrogate model consisting of $|K|$ components $z_k$, we define $U \in \mathbb{R}^{N\times K}$ where $u_{n k} = \cos \angle z_k n$ and $\mathbf{v} \in \mathbb{R}^N$ where $\mathbf{v} = \sum_{k=1}^K \mathfrak{s}_n \left(z_k\right)$.
For some linear signal representation $h\colon \mathbb{R}^N \to \mathbb{R}^M$, such as the identity mapping or a projection into a Fourier basis, the amplitude estimate is given by the ordinary least squares solution:

\begin{equation}
    \mathbf{\alpha}^\ast =
    \left(H(U)^T H(U)\right)^{-1} H(U)^T h\left(\mathbf{v}\right),
\end{equation}

\noindent where $H\colon \mathbb{R}^{N \times K} \to \mathbb{R}^{M \times K}$ applies $h$ to each column of a matrix. % i.e. $H(X) = \begin{bmatrix} h\left( \mathbf{x}_{\colon 1}\right) & \dots & h\left( \mathbf{x}_{ \colon k} \right)\end{bmatrix}$.
In practice, we may wish to select a nonlinear $h$ (e.g. the modulus of the DFT), leading to a nonlinear least squares problem.
However, when the nonlinear solution is reasonably well approximated by the linear solution for values of $|z_k^\ast|$ close to $1$, jointly learning the amplitude factor and multiplying $\hat{\alpha}_k\alpha_k^\ast$ appears to yield acceptable estimates.

\vspace{-1mm}
\section{Evaluation}
\label{sec:Evaluation}
\begin{figure}
    \centering
    \includegraphics[width=\columnwidth]{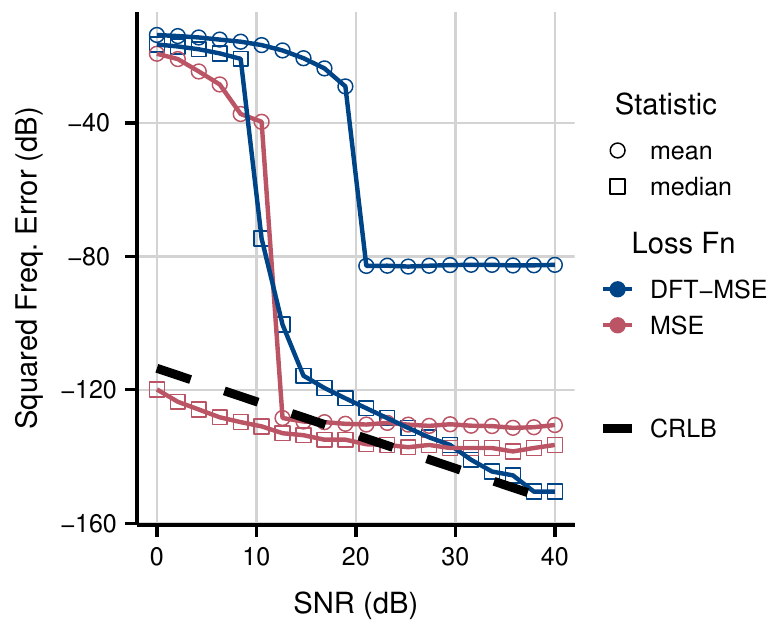}
    \caption{The results of sinusoidal frequency estimation by gradient descent for a single sinusoid in Gaussian white noise, using both DFT magnitude mean squared error (DFT-MSE) and time-domain mean squared error (MSE). Both mean and median squared error are plotted for each experiment. Also plotted for reference is the Cramér-Rao lower bound (CRLB) on unbiased frequency estimators.}
    \label{fig:single_freq}
\end{figure}

The performance of the surrogate model as a sinusoidal parameter estimator was evaluated by fitting the model to sinusoidal signals by gradient descent. All signals were of length $N=4096$.

\vspace{-1mm}
\subsection{Single sinusoid frequency estimation}\label{sec:single_freq}

In the single sinusoid case, we generated target signals with fixed amplitude $\alpha=1$ and initial phase $\phi=0$. The frequency parameter was sampled at 100 equal steps in the interval $[0.1\pi, 0.9\pi]$, and the signal-to-noise ratio (SNR) at 20 steps in the interval $[0, 40]$ dB, for a total of 2000 targets.
The corresponding signals were synthesized using the real valued sinusoidal model in Eqn \ref{eqn:model}.
A single starting parameter estimate was uniformly sampled from within the unit circle used for all 2000 targets, and the procedure was repeated with 10 different pseudo-random number generator seeds.
Optimization proceeded for 50k steps using the Adam optimizer with a learning rate of $0.0001$ and the mean squared error loss on either the time-domain signal or DFT magnitude spectrum.

Fig. \ref{fig:single_freq} displays the results of this experiment.
The mean and median squared error between the predicted and ground truth frequency parameters are plotted on a decibel scale (i.e. $10\log_{10}(\text{MSE})$).
The dotted black line plots the Cramér-Rao lower bound (CRLB) on variance for an unbiased estimator of sinusoidal frequency in Gaussian white noise, as given by Kay \cite{kay_fundamentals_1993}.
Whilst enquiry into the convergence properties of our method -- and therefore the underlying bias of the estimator -- is beyond the scope of this paper, this bound is representative of the performance of other sinusoidal parameter estimation algorithms.
We thus plot it here to facilitate comparison and to illustrate that the surrogate model with time domain MSE loss is capable of achieving an error comparable with non-gradient based estimators.

We note that the mean squared error of the frequency domain loss (DFT-MSE) does not fall below roughly $-83$dB, but the median squared error continues to fall, implying an increasingly skewed error distribution as the SNR rises.
We speculate that this occurs due to the loss of phase information in taking the modulus of the spectrum, and will investigate this hypothesis in future work.

% This evaluation was performed for both the single ($K=1$) and multiple ($K\in\left\{2,8,32\right\}$) component settings with a signal length of 4096.
% Estimation was performed using both the time-domain mean squared error loss -- with amplitude, frequency, and phase being estimated -- and DFT magnitude spectrum mean squared error loss -- predicting just amplitude and frequency.
% Descent was performed using the Adam optimiser with an initial learning rate of $0.0001$.

\vspace{-1mm}
\subsection{Multi-sinusoid frequency and amplitude estimation}\label{sec:multi_freq}

\begin{figure}[t]
    \centering
    \includegraphics[width=\columnwidth]{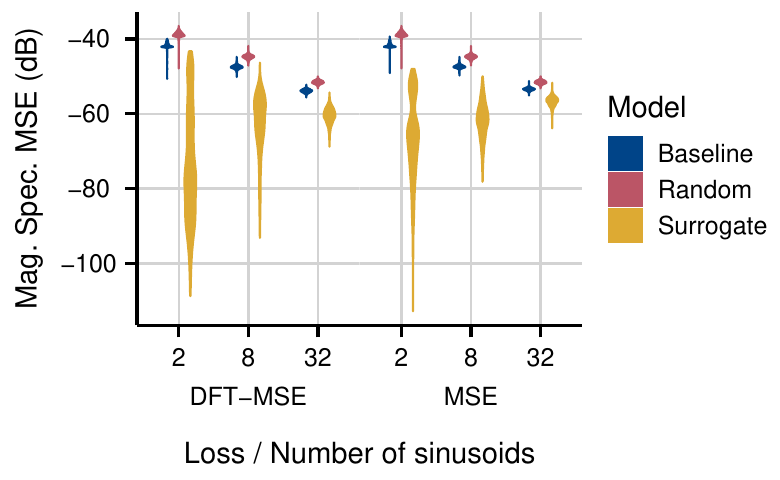}
    \caption{Violin plots of the results of sinusoidal frequency and amplitude estimation by gradient descent for mixtures of $|K|\in\{2, 8, 32\}$ sinusoids, using both DFT magnitude (DFT-MSE) and time-domain (MSE) mean squared error losses. Plots represent the distribution of mean squared errors between magnitude spectra across 2000 runs.}
    \label{fig:multi_freq}
\end{figure}

In the multi-sinusoid case, targets were generated with phase $\phi_k=0$, and frequency and amplitude sampled from uniform distributions, $\omega_k \sim \mathcal{U}(0.1\pi, 0.9\pi)$ and $\alpha_k \sim \mathcal{U}(0.1, 1.0)$.
2000 sets of target parameters were sampled, and the corresponding signals synthesized using the real valued sinusoidal model.
A random set of starting surrogate parameter estimates was uniformly sampled from the unit circle for each target, and linear amplitude was initialized for each component at $\hat{\alpha}_k = \frac{1}{|K|}$.
The experiment was repeated for $|K| \in \left\{2, 8, 32\right\}$.
Optimization ran for 100k steps using the same optimizer and losses.
As a baseline, the same procedure was applied, using the same targets and starting estimates, to a differentiable real valued sinusoidal model.

\begin{figure}[t]
    \centering
    \includegraphics[width=\columnwidth]{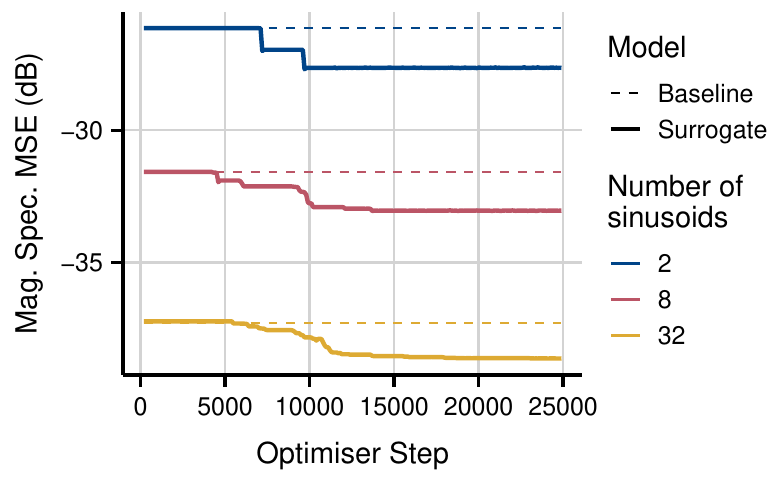}
    \caption{The mean squared error between estimated and target magnitude spectra, evaluated at every 100th optimizer step for baseline and surrogate models trained with time-domain MSE loss with $|K|\in\{2,8,32\}$.}
    \label{fig:descent_log}
\end{figure}

Fig. \ref{fig:multi_freq} displays the results of the experiment described in section \ref{sec:multi_freq}.
We plot the distributions of mean squared errors between target and predicted magnitude spectra using a dB scale for both our surrogate model and the real sinusoidal model baseline.
For comparison, we also plot the errors achieved with randomly sampled sinusoidal parameters.
Here, as expected, the surrogate clearly achieves superior performance, outperforming the baseline in all configurations.

We note that the performances of both the baseline and randomly sampled parameters improve as the number of components increases.
We speculate that this effect is due to the proportionally smaller expected distance between each model component and any target component for higher values of $|K|$.
Indeed, the decrease observed in the metric for both the baseline and random models is almost exactly proportional to the increase in the number of components -- that is, on the decibel scale we observe a change of $10\log_{10}\frac{1}{4}\approx -6.02$ for a $4\times$ increase in components.
%TODO : point out that the difference is 10*log(2/8) = 10*log(8/32) = -6.02

Conversely, the surrogate model's performance slightly degrades as $|K|$ increases.
Through informal observation of converged models, we hypothesize that this occurs due to a greater number of a specific class of local minimum, wherein multiple model components combine to match a single component in the target signal.
We leave formal study of this behaviour to future work, but note that this phenomenon seems to predominantly occur only at the expense of quieter signal components.

To illustrate the optimization dynamics of the surrogate, Fig. \ref{fig:descent_log} plots the evolution of the metric throughout optimization for both the surrogate and baseline model for a randomly selected set of target parameters.
Here we see that the baseline metric either does not fall, or falls imperceptibly, as should be expected given the properties described in Section \ref{sec:sinusoidal}.
The surrogate metric, however, does clearly fall before converging on a final value.
It appears to solve the multi-sinusoid problem sequentially -- that is, it seems to resolve each component one-by-one, causing the metric to fall to a series of plateaus.
This observation may have implications for training strategies in DDSP deep learning tasks, where a plateau in a metric is typically taken as a signifier that a model has converged.

\vspace{-1mm}
\section{Conclusion}
\label{sec:conclusion}

This work presented a technique for matching the frequency and amplitude parameters of a single- and multi-component sinusoidal model to a target signal by gradient descent.
We evaluated the performance of our method on single and multiple sinusoid signals and demonstrated that it clearly outperforms a standard sinusoidal model in the multi-sinusoid case, whilst approaching the performance of other, non-gradient based estimators in the single sinusoid case.

This problem was previously intractable using differentiable signal processing techniques, preventing a variety of applications of this family of methods, including the modelling of inharmonic audio signals, unsupervised fundamental frequency detection, and more.
Our approach now paves the way for these applications to be explored. 
In particular, we believe our surrogate model is suitable for use as a drop-in replacement for a differentiable sinusoidal model, and in future work will explore its capabilities in end-to-end learning with differentiable signal processing.
We will also conduct further study into the surrogate model's optimization characteristics.

% \small
\bibliographystyle{IEEEbib}
\bibliography{references,strings,refs}

\end{document}